\documentclass[
    superscriptaddress,
    reprint,
    showpacs,
    amsmath,
    amssymb,
    aps,
    prb,
    floatfix
]{revtex4-1}
\usepackage{graphicx}
\usepackage{dcolumn}
\usepackage{bm}
\usepackage{braket}
\usepackage{subfigure}
\usepackage{hyperref}
\hypersetup{
    colorlinks=true,
    linkcolor=blue,
    citecolor=blue,
    urlcolor=blue
}
\begin{document}
\title{
Effect of doping on lattice instabilities of single-layer 1H-TaS$_2$
}
\author{Oliver R. Albertini}
\email{ora@georgetown.edu}
\affiliation{
    Department of Physics,
    Georgetown University,
    Washington, DC 20057, USA
}
\author{Amy Y. Liu}
\affiliation{
    Department of Physics,
    Georgetown University,
    Washington, DC 20057, USA
}
\author{Matteo Calandra}
\affiliation{
IMPMC, CNRS, Universit\'e P. et M. Curie, 4 Place Jussieu,
75005 Paris, France
}
\date{\today}
\begin{abstract}
Recent ARPES measurements of single-layer 1H-TaS$_2$ grown on Au(111) suggest strong electron doping from the substrate.
In addition, STM/STS measurements on this system show suppression of the charge-density-wave (CDW) instability that occurs in bulk 2H-TaS$_2$.
We present results from {\it ab initio} DFT calculations of free-standing single-layer 1H-TaS$_2$ to explore the effects of doping on the CDW.
In the harmonic approximation,  we find that a lattice instability along the $\Gamma M$ line occurs in the undoped monolayer, consistent with the bulk $3\times3$ CDW ordering vector.
Doping removes the CDW instability, in agreement with the experimental findings.
The doping and momentum dependence of both  the electron-phonon coupling and of the bare phonon energy (unscreened by metallic electrons) determine the stability of lattice vibrations.
Electron doping also causes an expansion of the lattice, so strain is a secondary but also relevant effect.
\end{abstract}

\maketitle

\section{\label{sec:intro}Introduction}
A charge-density wave (CDW) is a common collective phenomenon in solids consisting of a periodic modulation of the electron density accompanied by a distortion of the crystal lattice.
For a one-dimensional system of non-interacting electrons, the phenomenon is described by the Peierl's instability, where a divergence in the real part of the electronic susceptibility at twice the Fermi wave vector drives a metal-to-insulator transition.
In extensions of this idea to real materials with anisotropic band structures and quasi-1D Fermi surfaces, Fermi-surface nesting is often cited to explain charge-ordering tendencies.
However, geometric Fermi-surface nesting (i.e., existence of parallel regions of the Fermi surface separated by a single wave vector {\bf q}) is related to the {\em imaginary} part of the noninteracting susceptibility rather than the real part, so it is not directly connected to the Peierl's mechanism.\cite{johannes-mazin}
Indeed, calculations for a number of CDW materials, including NbSe$_2$, TaSe$_2$, TaS$_2$, and CeTe$_3$, have found little or no correlation between the CDW ordering vector {\bf q}$_{\mathrm{CDW}}$, and peaks in the geometric nesting function (imaginary part of $\chi_0$).\cite{johannes-mazin, johannes-mazin-howells, calandra-mazin-mauri, ge-liu, liu-TaS2}
Nor does {\bf q}$_{\mathrm{CDW}}$ coincide with sharp features in the real part of $\chi_0$ in most of these cases.
Instead, these studies pointed to the importance of the momentum-dependent electron-phonon coupling, which softens selected phonon modes to the point of instability.

Advances in the synthesis of low-dimensional materials provide new opportunities to test these ideas.
In going from quasi-2D bulk NbSe$_2$ to the 2D monolayer, for example, the number of bands crossing the Fermi level decreases from three to one, which could reveal information about the role played by different bands in driving the CDW instability.
The stability and structure of the CDW phase in monolayer NbSe$_2$, however, remains under debate.
Density functional calculations predict that both the monolayer and bilayer have CDW instabilities, but with a shifted {\bf q}$_{\mathrm{CDW}}$ and a larger energy gain compared to the bulk.\cite{calandra-mazin-mauri}
A recent experimental study of bulk and 2D NbSe$_2$ on SiO$_2$ used the Raman signature of the CDW amplitude mode to estimate the CDW transition temperature and reported an increase of $T_{\mathrm{CDW}}$ from about 33 K in the bulk to 145 K in the monolayer.\cite{mak}
This result, attributed in Ref. \onlinecite{mak} to an enhanced electron-phonon coupling due to reduced screening in two dimensions, is qualitatively consistent with the DFT predictions, but does not address the question of the CDW superstructure.
On the other hand, another study of single-layer NbSe$_2$, this time on bilayer graphene, reported scanning tunneling microscopy/spectroscopy (STM/STS) measurements showing a CDW transition at a lower temperature than the bulk but with the same {\bf q}$_{\mathrm{CDW}}$ ordering vector.\cite{crommie}
Discrepancies between the two experimental studies and between experiment and theory could be due in part to substrate effects, as 2D materials tend to be highly sensitive to the environment and to the substrate.
Initial studies of 1T-TaS$_2$, for example, suggested that it completely loses the low-temperature commensurate CDW phase upon thinning to about 10-15 layers,\cite{yu} but later it was shown that oxidation can suppress the formation of the commensurate phase.\cite{tsen}
Raman signatures of the commensurate CDW were later found in single-layer 1T-TaS$_2$ samples with limited exposure to air.\cite{albertini}
In using 2D materials to probe the CDW transition and the effect of dimensionality, it is therefore crucial to differentiate behavior intrinsic to each material from effects due to the environment.

In bulk form, TaS$_2$ is a layered transition-metal dichalcogenide for which the 1T and 2H polymorphs are competitive in energy, and both can be synthesized.
Both the 1T and 2H polymorphs undergo CDW transitions as the temperature is lowered.
Recently, it was reported that single-layer TaS$_2$ grown epitaxially on a Au (111) substrate adopts the 1H structure rather than 1T.\cite{sanders}
Low-energy electron diffraction (LEED) and STM data indicate that on Au (111), single-layer 1H-TaS$_2$ does not undergo a CDW transition, at least down to $T = 4.7$ K.
This is in contrast to bulk 2H-TaS$_2$, which develops a $3\times3$ CDW periodicity below about 75 K.
In addition, a comparison of angle-resolved photoemission spectroscopy (ARPES) data to the band structure calculated for a free-standing monolayer suggests that the substrate causes the material to become $n$-doped with a carrier concentration of approximately 0.3 electrons per unit cell.\cite{sanders}

In this paper, we use DFT calculations to investigate whether the observed suppression of the CDW in monolayer 1H-TaS$_2$ is intrinsic or a consequence of its interaction with the metallic substrate.
In the harmonic approximation, we find that a free-standing single layer of 1H-TaS$_2$ has lattice instabilities that are very similar to its bulk counterpart.
When the monolayer is electron-doped, however, the 1H structure becomes dynamically stable, implying that substrate-induced doping is the likely reason for the suppression of the CDW observed in Ref. \onlinecite{sanders}.
The strong effect of doping and a secondary effect of lattice strain on the CDW transition in this material give insight on what drives the transition.
We examine the role of electron-phonon coupling by calculating the phonon self-energy.
We find that the contribution to the real part of the phonon self-energy from the single band that crosses the Fermi level is largely responsible for the momentum dependence of the soft mode, but the bare phonon frequency (unscreened from electrons at the Fermi level) also plays a role.

\section{\label{sec:methods}Methods}
We performed density-functional theory calculations using the QUANTUM-ESPRESSO\cite{espresso} package.
The exchange-correlation interaction was treated with the Perdew-Burke-Ernzerhof (PBE) generalized gradient approximation.\cite{pbe}
To describe the interaction between electrons and ionic cores, we used ultrasoft\cite{uspp} Ta pseudopotentials and norm-conserving\cite{NCpseudo} S pseudopotentials.
Electronic wave functions were expanded in a plane-wave basis with kinetic energy cutoffs of 47 (53) Ry for scalar (fully) relativistic pseudopotentials.
Integrations over the Brillouin zone (BZ) were performed using a uniform grid of $36\times36\times1$ {\bf k} points, with an occupational smearing width of $\sigma = 0.005$ Ry.

To simulate electronic doping, electrons were added to the unit cell along with a compensating uniform positive background.
A single layer of 1H-TaS$_2$ was modeled using a supercell with the out-of-plane lattice parameter fixed at $c = 12$ \AA, corresponding to about 9 \AA~ of vacuum between layers.
This ensured that for the range of electron doping explored, spurious vacuum states that originate from the periodic boundary conditions stay well above the Fermi level.
The in-plane lattice constant and the atomic positions were fully relaxed at each doping level.

Phonon spectra and electron-phonon matrix elements were calculated with density functional perturbation theory\cite{dfpt} on a {\bf q}-point grid of $12\times12\times1$ phonon wave vectors and Fourier interpolated to denser grids.
To test the effect of spin-orbit coupling on the phonon frequencies, we used a less dense grid of $8\times8\times1$ {\bf q}-points.
While this does not capture all the fine structure in the phonon dispersions, it is sufficient to test the effect of spin-orbit coupling.
In fact, even Fourier interpolation on the $12\times12\times1$ {\bf q}-grid does not fully capture the sharp features along certain directions in the Brillouin zone where direct calculations of the phonon frequency are needed.
The real and imaginary parts of electronic susceptibilities and phonon self energies were calculated using dense {\bf k}-point grids of at least $72\times72\times1$ points.

\section{\label{sec:results}Results \& Discussion}
The 1H-TaS$_2$ lattice has point group $D_{3h}$, reduced from $D_{6h}$ in the bulk.
Sulfur atoms adopt a trigonal prismatic coordination about each Ta site, and the crystal lacks inversion symmetry.
Our calculations for the undoped ($x=0$) monolayer give a relaxed in-plane lattice parameter of $a_{x=0} = 3.33$ \AA~and a separation between Ta and S planes of S$_z = 1.56$ \AA.
These values are similar to what we calculate for the bulk.
For comparison, the experimental in-plane lattice parameter for single-layer 1H-TaS$_2$ on Au\cite{sanders} and bulk 2H-TaS$_2$\cite{conroy-pisharody} are $a = 3.3(1)$ \AA~and 3.316 \AA, respectively.
With the addition of 0.3 electrons per cell ($x = -0.3$), the lattice expands roughly 2.5\% to $a_{x=-0.3} = 3.41$ \AA, while the separation between Ta and S planes decreases slightly to S$_z = 1.53$ \AA.
In analyzing the impact of doping on the CDW instability, we examine the effect of adding charge carriers as well as the effect of the lattice expansion.

\begin{figure}
   \includegraphics[width=0.85\linewidth]{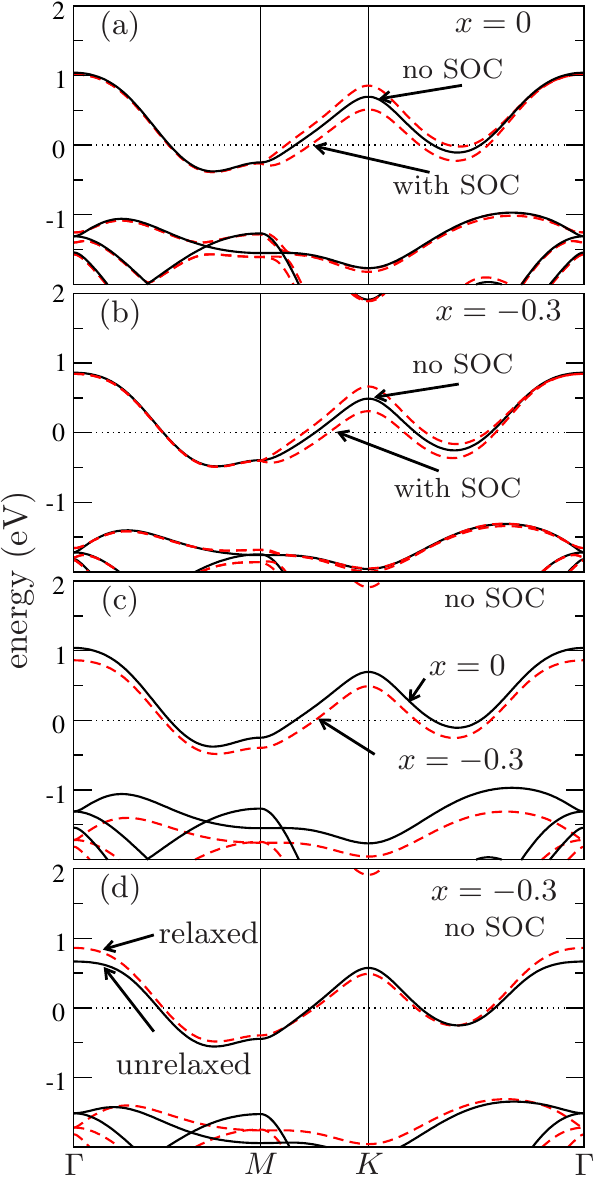}
   \caption
   {
      Electronic bands of single-layer 1H-TaS$_2$.
      Effect of:
      (a) SOC for the undoped case,
      (b) SOC for the $n$-doped ($x = -0.3$) case,
      (c) $n$-doping ($x = -0.3$ vs. $x = 0$),
      (d) lattice constant relaxation ($a_{x=0}$ vs. $a_{x=-0.3}$) on the doped ($x=-0.3$) system.
      Internal atomic coordinates were optimized in all cases.
   }
   \label{fig:bands}
\end{figure}

In single-layer 1H-TaS$_2$, in the absence of spin-orbit coupling (SOC), a single isolated band (two-fold spin degenerate) crosses the Fermi level, as shown in Fig. \ref{fig:bands}(a).
This half-filled band, which is about 1.4 eV wide, has strong Ta $d$ character and is separated by about 0.6 eV from the manifold of S $p$ bands below.
As a consequence, metallic screening is due to a single band isolated from all the others.
The Fermi surface consists of a roughly hexagonal hole sheet of primarily Ta $d_{z^2}$ character around $\Gamma$ and a roughly triangular sheet of primarily in-plane Ta $d$ character ($d_{x^2+y^2}$ and $d_{xy}$) around $K$.
The spin-orbit interaction splits the half-filled $d$ band by as much as $\sim$~0.3 eV, except along the $\Gamma M$ line where symmetry dictates that the band remains degenerate.
The number of Fermi sheets around the $\Gamma$ and $K$ points doubles with SOC.
\begin{figure}
   \includegraphics[width=0.85\linewidth]{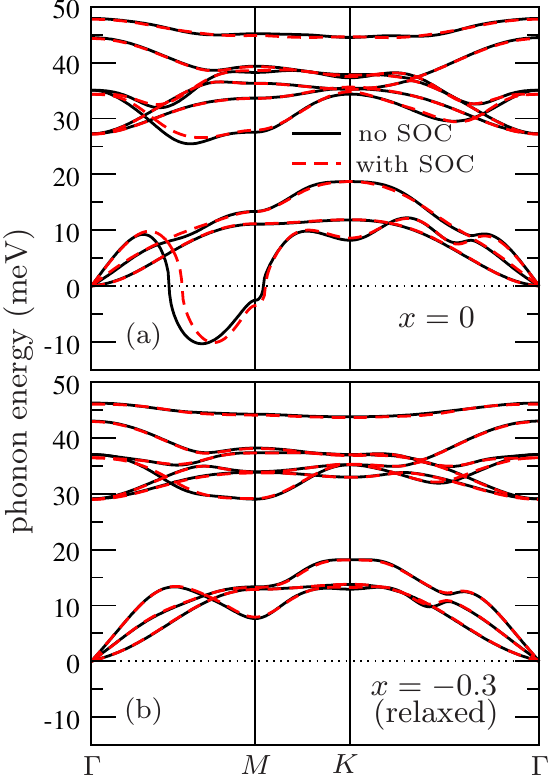}
   \caption
   {
      Phonon dispersions of single-layer 1H-TaS$_2$.
      Effect of SOC for:
      (a) the undoped case,
      (b) the $n$-doped ($x = -0.3$) case.
      In (b), the curves with and without SOC lie almost directly on top of each other.
      Dispersion curves were obtained from Fourier interpolation from a {\bf q}-grid of $8\times8\times1$ points.
      Imaginary frequencies are shown as negative.
   }
   \label{fig:phononsSOC}
\end{figure}

Electron doping of $x=-0.3$ causes a nearly rigid downward shift of the partially occupied Ta $d$ band by about 0.1 eV, and a somewhat larger downward shift of the occupied S $p$ bands.
This is shown in Fig. \ref{fig:bands}(c) for the scalar relativistic bands. 
The fully relativistic bands behave similarly with doping, as can be seen in Figs. \ref{fig:bands}(a,b).
The hole pocket around $\Gamma$ thus gets slightly smaller while the one around $K$ shrinks more significantly.
The doping-induced lattice expansion, on the other hand, has a negligible effect on the Fermi surface, as shown in Fig. \ref{fig:bands}(d).

Despite the non-negligible effect that SOC has on the electronic structure, we find little difference between the phonon spectra calculated with and without SOC.
This is shown in Fig. \ref{fig:phononsSOC} where the phonon dispersion curves were obtained from a Fourier interpolation on a $8\times8\times1$ phonon {\bf q}-grid.
Hence, for the remainder of this paper, we focus on the phonon properties calculated in the absence of SOC.

\begin{figure}
   \includegraphics[width=0.85\linewidth]{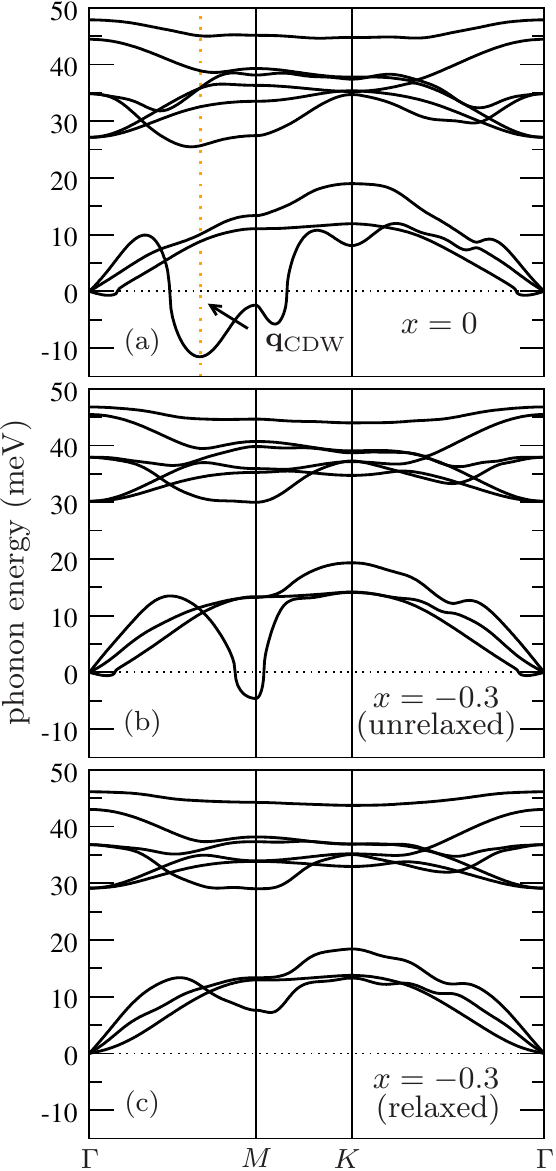}
   \caption
   {
      Phonon dispersions of single-layer 1H-TaS$_2$:
      (a) undoped ($x=0$),
      (b) $n$-doped with unrelaxed lattice constant ($x=-0.3$, $a_{x=0}$),
      (c) $n$-doped with relaxed lattice constant ($x=-0.3$, $a_{x=-0.3}$).
      Dispersion curves were obtained from Fourier interpolation from a {\bf q}-grid of $12\times12\times1$ points.
      Imaginary frequencies are shown as negative.
   }
   \label{fig:phonons}
\end{figure}

\begin{figure*}
   \includegraphics[width=0.85\textwidth]{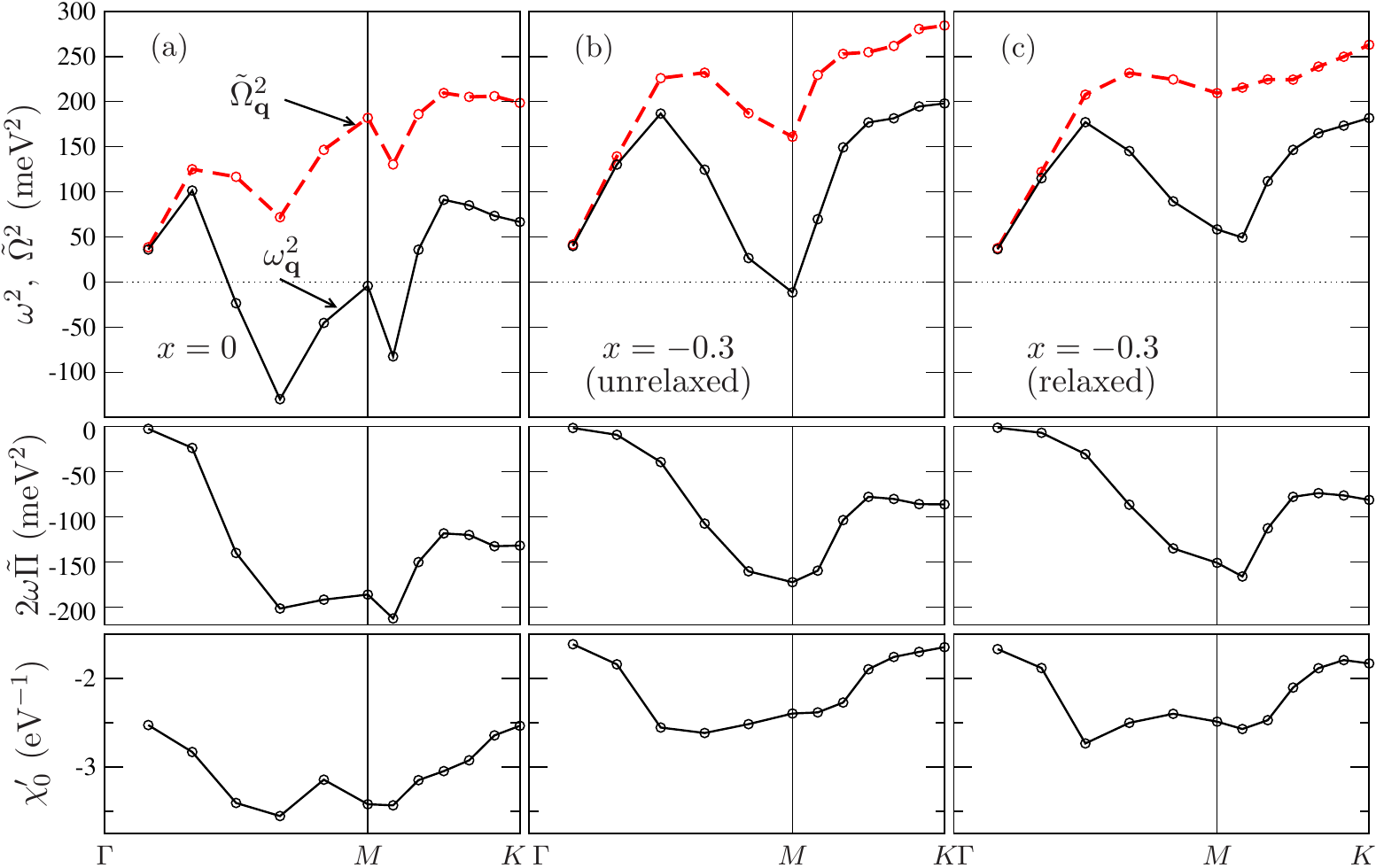}
   \caption
   {
      Momentum dependence of the square of the self-consistent ($\omega_{\bf q}^2$) and bare ($\tilde{\Omega}_{\bf q}^2$) phonon energies for the branch with instabilities, self-energy correction ($2\omega_{\bf q}\tilde\Pi_{\bf q}$) for that branch, and real part of the bare susceptibility ($\chi_{0}'({\bf q})$):
      (a) undoped ($x=0$),
      (b) $n$-doped with unrelaxed lattice constant ($x=-0.3$, $a_{x=0}$),
      (c) $n$-doped with relaxed lattice constant ($x=-0.3$, $a_{x=-0.3}$).
      The red, dashed lines in the top panels represent the square of the bare phonon energies $(\tilde{\Omega}_{{\bf q}}^2)$, calculated using Eq. \ref{eq:perturbation-theory}.
      Only the response of the band crossing the Fermi level is included in $2\omega_{\bf q}\tilde\Pi_{\bf q}$ and $\chi_{0}'({\bf q})$.
   }
   \label{fig:self-energy}
\end{figure*}

Figure \ref{fig:phonons}(a) shows the phonon dispersion curves calculated for the undoped monolayer ($x = 0$) from a Fourier interpolation on a $12\times12\times1$ phonon {\bf q}-grid.
An acoustic phonon branch is found to be unstable over a large area of the BZ surrounding the $M$ point and even has a dip at $K$.
(Imaginary frequencies are plotted as negative.)
This branch involves in-plane Ta vibrations.
This region of instability arises primarily from softening of the phonon branch near the bulk CDW wavevector (2/3 along the $\Gamma$ to $M$ line), but there is also a secondary point of instability along the $M$ to $K$ line (close to $M$).
\footnote{
   While the instability at ${\bf q}_{MK}$ is not captured by Fourier interpolation of the $8\times8\times1$ {\bf q}-grid in Fig. \ref{fig:phononsSOC}(a), the Fourier interpolation based on a $12\times12\times1$ {\bf q}-grid in Fig. \ref{fig:phonons}(a) clearly shows the instability.
   Even so, direct calculation at ${\bf q}_{MK}$ yields an even more imaginary phonon frequency [Fig. \ref{fig:self-energy}(a)].
}
We will refer to these points as ${\bf q}_{\mathrm{CDW}}$ and ${\bf q}_{MK}$, respectively.

When electrons are added while keeping the lattice constant fixed at $a_{x=0}$ but allowing S$_z$ to relax, the instabilities at ${\bf q}_{\mathrm{CDW}}$ and ${\bf q}_{MK}$ are progressively suppressed.
With doping of $x=-0.3$, a weak instability at the $M$ point remains, as shown in Fig. \ref{fig:phonons}(b).
Once the lattice constant is allowed to expand to its optimal value at $x=-0.3$, the lattice becomes dynamically stable, as seen in Fig. \ref{fig:phonons}(c), though an anomalous dip in the phonon dispersion remains near ${\bf q}_{MK}$.
For both doped cases, there is no longer a dip at $K$.

Since the unstable branch involves in-plane displacements of Ta ions, it is likely that these phonons couple strongly to in-plane Ta $d$ states near the Fermi level.
To investigate the role of the electron-phonon interaction in the CDW transition in this material, we consider the phonon self-energy due to electron-phonon coupling.
In the static limit, the real part of the phonon self-energy for phonon wave vector {\bf q} and branch $\nu$ is given by
\begin{equation}
   \Pi_{{\bf q}\nu} =\frac{2}{N_{\bf k}}\sum_{{\bf k}jj'}|g^{\nu}_{{\bf k}j,{\bf k} + {\bf q}j'}|^2\frac{f(\epsilon_{{\bf k}+{\bf q}j'})-f(\epsilon_{{\bf k}j})}{\epsilon_{{\bf k}+{\bf q}j'}-\epsilon_{{\bf k}j}}, \label{eq:se-real}\\
\end{equation}
while the phonon linewidth, which is twice the imaginary part of the phonon self-energy, can be expressed as
\begin{equation}
   \gamma_{{\bf q}\nu} = \frac{4\pi\omega_{{\bf q}\nu}}{N_{\bf k}}\sum_{{\bf k}jj'}|g^{\nu}_{{\bf k}j,{\bf k} + {\bf q}j'}|^2\delta(\epsilon_{{\bf k}j} - \epsilon_F)\delta(\epsilon_{{\bf k} + {\bf q}j'} - \epsilon_F). \label{eq:se-imag}
\end{equation}
Here $N_{\bf k}$ is the number of {\bf k} points in the Brillouin zone, $j$ and $j'$ are electronic band indices, $f(\epsilon)$ is the Fermi-Dirac function, $\epsilon_F$ is the Fermi energy, and $\omega_{{\bf q}\nu}$ is the phonon frequency.
The electron-phonon matrix element is
\begin{equation}
\label{eq:epme}
g_{{\bf k}j,{\bf k} + {\bf q}j'}^{\nu} = \sqrt{\frac{\hbar}{2\omega_{{\bf q}\nu}}}\bra{{\bf k}j}\delta V_{\mathrm{SCF}}/\delta u_{{\bf q}\nu} \ket{{\bf k} + {\bf q}j'},
\end{equation}
where $u_{{\bf q}\nu}$ is the amplitude of the phonon displacement and $V_{\mathrm{SCF}}$ is the Kohn-Sham potential.
Both the linewidth and the product $\omega_{{\bf q}\nu}\Pi_{{\bf q}\nu}$ are independent of $\omega_{{\bf q}\nu}$, so they remain well defined even when the frequency is imaginary.

\begin{figure}
   \includegraphics[width=0.90\linewidth]{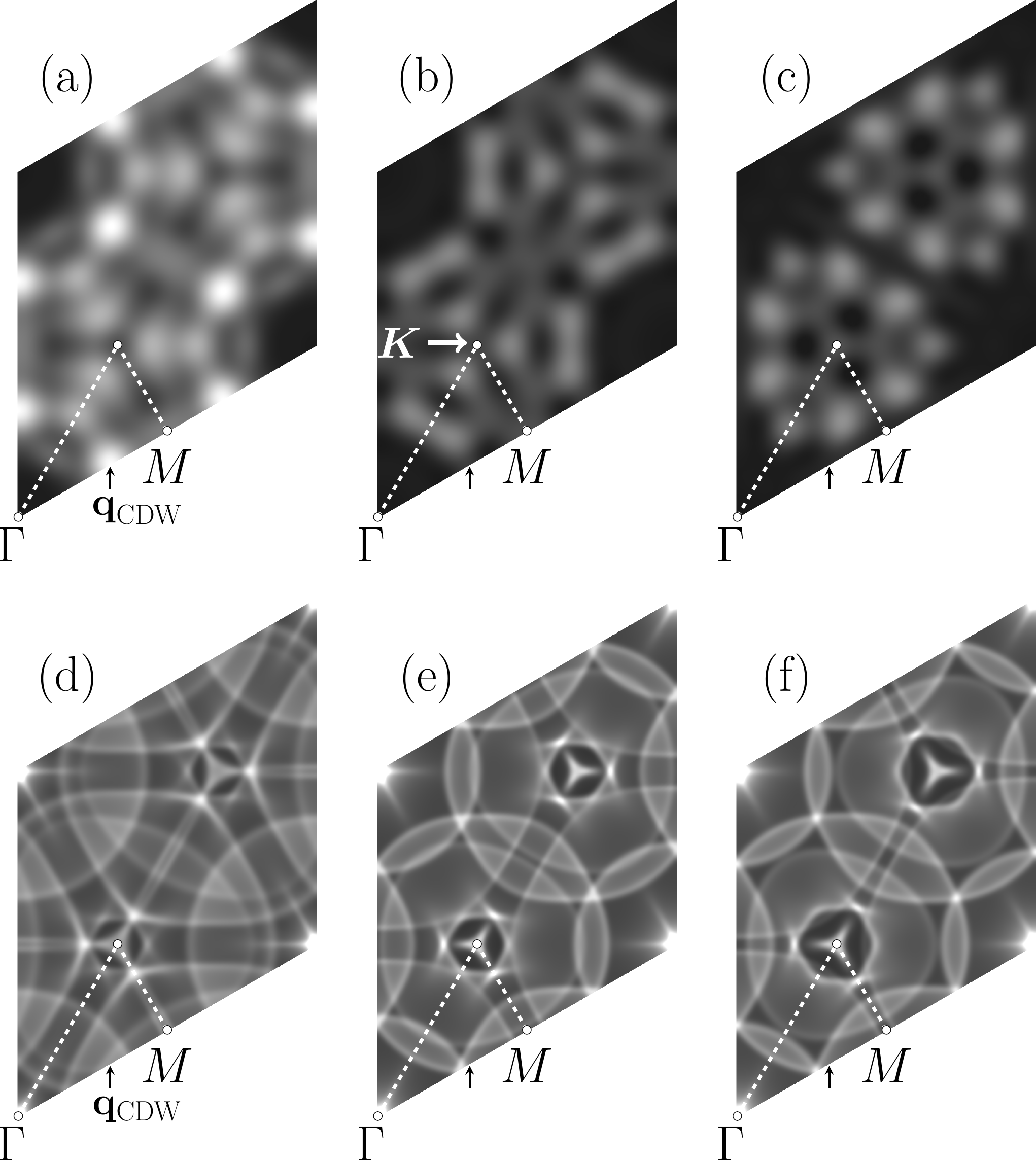}
   \caption
   {
      Brillouin-zone maps of the phonon linewidth [(a-c)] and the geometric Fermi-surface nesting function [(d-f)]:
      (a,d) undoped ($x=0$),
      (b,e) $n$-doped with unrelaxed lattice constant ($x=-0.3$, $a_{x=0}$),
      (c,f) $n$-doped with relaxed lattice constant ($x=-0.3$, $a_{x=-0.3}$).
      White (black) represents high (low) values, with one scale used for all the linewidth plots (a-c) and another used for all the nesting function plots (d-f).
      Dotted lines show high-symmetry lines from $\Gamma$ to $K$ and from $K$ to $M$.
      The plots in (a-c) represent the sum of the phonon linewidths for the two acoustic phonon branches with in-plane Ta displacements.
   }
   \label{fig:gamma-chi}
\end{figure}

In Fig. \ref{fig:gamma-chi}, we show Brillouin zone maps of the phonon linewidth (summed over the two acoustic modes with in-plane Ta displacements).
In the undoped material [Fig. \ref{fig:gamma-chi}(a)], the linewidth is sharply peaked at ${\bf q}_{\mathrm{CDW}}$.
For comparison, the geometric nesting function, which is defined as the Fermi surface sum in Eq. \ref{eq:se-imag} with constant matrix elements, does not have sharp structure near ${\bf q}_{\mathrm{CDW}}$ [Fig. \ref{fig:gamma-chi}(d)], and instead has three sharp peaks surrounding the $K$ point.
This means that the sharp peak in the linewidth at ${\bf q}_{\mathrm{CDW}}$ must be due to large electron-phonon matrix elements between states on the Fermi surface, rather than to the geometry of the Fermi surface itself.
With electron doping ($x=-0.3$), the maximum values of the linewidth are much smaller and occur elsewhere in the Brillouin zone, whether the lattice constant is allowed to relax [Fig. \ref{fig:gamma-chi}(c)] or not [Fig. \ref{fig:gamma-chi}(b)].
This suggests that the momentum dependence of the electron-phonon coupling of states at the Fermi level plays an important role in picking out the primary instability at ${\bf q}_{\mathrm{CDW}}$.
However, focusing on states at the Fermi level is not sufficient to understand the broader region of instability.

The real part of the phonon self-energy $\Pi_{{\bf q}\nu}$ is directly related to the renormalization of phonon frequencies due to electron screening.
From perturbation theory,
\begin{equation}
   \label{eq:perturbation-theory}
   \omega_{{\bf q}\nu}^2 =\Omega_{{\bf q}\nu}^2 + 2 \omega_{{\bf q}\nu} \Pi_{{\bf q}\nu},
\end{equation}
where $\Omega_{{\bf q}\nu}$ is the bare phonon frequency.
While it is conceptually appealing to define the bare frequencies as the completely unscreened ionic frequencies, such a starting point leads to results that lie outside the range of validity of the perturbative expression in Eq. \ref{eq:perturbation-theory}.
Instead, we define the bare frequencies to be the frequencies obtained neglecting the metallic screening due to electrons in the isolated Ta $d$ band crossing the Fermi level.
It has been shown that this non self-consistent definition is perfectly justified in the framework of linear response theory.\cite{calandra-profeta-mauri}
In this case, we limit the sum in the phonon self-energy (Eq. \ref{eq:se-real}) to intraband transitions within that band and denote it as $\tilde{\Pi}_{{\bf q}\nu}$.
The corresponding bare (in the sense of unscreened by metallic electrons) frequencies are denoted $\tilde{\Omega}_{{\bf q}\nu}$.
While the bare frequencies are not directly accessible, we can estimate them from the fully screened frequencies $\omega_{{\bf q}\nu}$ (as obtained from density functional perturbation theory) and the perturbative correction $2 \omega_{{\bf q}\nu} \tilde{\Pi}_{{\bf q}\nu}$ by inverting Eq. \ref{eq:perturbation-theory}.

For the branch with instabilities, Fig. \ref{fig:self-energy} shows the square of the phonon frequencies and the self-energy correction, $2\omega_{{\bf q}\nu}\tilde{\Pi}_{{\bf q}\nu}$, along high-symmetry directions in the Brillouin zone.
For comparison, $\chi_0'({\bf q})$, the real part of the bare electronic susceptibility, given by Eq. \ref{eq:se-real} with constant matrix elements and limited to the band at the Fermi level, is also plotted.
For the undoped material, $\chi_0'$ has a minimum at ${\bf q}_{\mathrm{CDW}}$, but the momentum dependence between ${\bf q}_{\mathrm{CDW}}$ and $M$ does not match that of the calculated frequencies.
Similarly, the momentum dependence of $\chi_0'$ does not correlate with the phonon softening when the material is doped to $x=-0.3$, with or without relaxation of the lattice constant.
From Fig. \ref{fig:self-energy}, we see, however, that the momentum dependence of the phonon self-energy roughly follows that of the phonon softening, indicating the importance of the electron-phonon matrix elements in $\tilde\Pi_{{\bf q}\nu}$.
The self-energy includes contributions from states throughout the band crossing the Fermi level, in contrast to the linewidth, which provides information about the electron-phonon coupling at the Fermi level.

The square of the bare frequencies ($\tilde\Omega_{{\bf q}\nu}^2$) estimated using Eq. \ref{eq:perturbation-theory} are plotted with dashed lines in the top panels of Fig. \ref{fig:self-energy}.
Note that in the regions of instabilities, the self-energy correction is comparable to or larger than the square of the bare frequency, possibly raising into question the degree to which the perturbative expression is valid.
Nevertheless, if we use the expression to estimate the bare frequencies, we find some surprising results.
For the undoped material, the dispersion of the bare frequency has sharp dips at both ${\bf q}_{\mathrm{CDW}}$ and ${\bf q}_{MK}$.
Thus, while phonon softening due to screening of the band at the Fermi level accounts for most of the momentum dependence of the unstable modes, structure in the bare frequencies contributes as well.
Upon doping to $x=-0.3$, but holding the lattice constant fixed at $a_{x=0}$, the local minima at ${\bf q}_{\mathrm{CDW}}$ and ${\bf q}_{MK}$ disappear in both the self-energy and the bare frequency, stabilizing those modes.
At $M$, however, neither the self-energy nor the bare frequency change significantly with doping, so a residual instability remains at $M$.
When the lattice constant of the doped system is relaxed, the self-energy undergoes relatively minor changes, while the bare frequency at $M$ hardens significantly.
Thus it is the lattice-constant dependence of the bare frequency that stabilizes the $M$ point phonon at a doping of $x=-0.3$.
The dramatic hardening of this mode upon a 2.5\% lattice expansion is surprising, as phonons usually soften with lattice expansion.
Indeed most of the other phonon modes in the doped material soften slightly with the lattice expansion.

To summarize, we find that the primary CDW ordering vector coincides with very strong electron-phonon coupling at the Fermi surface, but the full momentum dependence of the phonons in the wide region of instability is described more completely by the momentum dependence of $\omega\tilde{\Pi}$, the phonon self-energy due to screening from the band crossing the Fermi level.
The momentum dependence of the corresponding bare phonon frequencies, $\tilde\Omega$, however, also plays a role.
Regarding the behavior of the instabilities with doping and lattice strain, the bare frequencies provide the primary contributions, but the changes in the phonon self-energy, especially with doping, are essential.
While recent studies of CDW materials have focused on the role of electron-phonon coupling as the driving mechanism, we find that, for monolayer 1H-TaS$_2$, taking into account screening arising from electron-phonon coupling in the band crossing the Fermi level by itself is not enough, and one must consider how the bare frequencies depend on momentum and doping as well.
Unfortunately it is not possible to determine if the momentum, doping, and strain dependencies deduced for the bare frequencies come from the response of other bands away from the Fermi level or appear as features in the completely unscreened ionic frequencies.

\section{\label{sec:conlusion}Conclusions}
Our calculations show that in the harmonic approximation the free-standing monolayer of 1H-TaS$_2$ is unstable to CDW distortions with the same ordering vector as the bulk.
We also find that electron doping stabilizes the lattice.
These results indicate that the CDW suppression found in experiments of 1H-TaS$_2$ on Au is not intrinsic but rather induced by the substrate.\cite{sanders}
According to our harmonic calculations, in addition to increasing the number of charge carriers, electron doping induces a lattice expansion.
While the addition of charge carriers itself stabilizes most of the soft phonon modes, at the harmonic level a residual instability remains if the lattice constant is not allowed to relax.
In the experiments on the Au (111) substrate, the uncertainty in the measured lattice constant was about 3\%,\cite{sanders} which is slightly larger than the lattice expansion we predict for electron doping of $x=-0.3$.
Hence in the experimental system, it remains an open question as to how much the substrate affects the lattice constant, and how that in turn influences the lattice instabilities.

We believe that the suppression of CDW by electron doping is robust also against anharmonic effects. Indeed in metals, anharmonic effects tend to enhance phonon frequencies and  suppress CDW instabilities. These effects, not considered in this work, could reduce the tendency towards CDW at $x=0$, but will not change the qualitative result that electron doping removes the CDW in this system.

Recently, it was pointed out that for metallic 2D materials on substrates, the shift in the bands measured in ARPES experiments may not provide a reliable estimate of charge transfer across the interface.\cite{wehling}
In particular, if there is significant hybridization between the substrate and 2D material, the actual charge transfer will be reduced.
The impact of substrate hybridization on the CDW instability in monolayer 1H-TaS$_2$ warrants future investigation.

This system, like other transition-metal dichalcogenides, offers an opportunity to better understand the intrinsic origins of CDWs.
Going beyond previous studies that emphasized the importance of the momentum dependence of the electron-phonon interaction in driving the CDW, we find that the momentum, doping, and strain dependencies of the bare phonon frequencies also play a role.
This system also underscores the importance of disentangling environmental and intrinsic effects in 2D materials, and provides an example of using the substrate to tune the properties of the material.

\begin{acknowledgments}
This work was supported by NSF Grant EFRI-143307.
M.C. acknowledges support from Agence Nationale de la Recherche under contract ANR-13-IS10-0003-01, from the Graphene Flagship, PRACE for awarding us access to resources on Marenostrum at BSC and the computer facilities provided by CINES, IDRIS, and CEA TGCC (Grant EDARI No.  2017091202).
\end{acknowledgments}

\end{document}